\documentclass[a4paper,11pt]{article}

\usepackage{graphicx}
\usepackage{color}
\usepackage[top=4cm, bottom=4cm, left=2.5cm, right=2.5cm]{geometry}
\usepackage{amsmath}
\usepackage{amssymb}
\usepackage{authblk}
\usepackage{float}
\usepackage{natbib}
\usepackage[margin=1cm,font=small,labelfont=bf]{caption}
\usepackage[usenames,dvipsnames]{xcolor}

\usepackage{fancyhdr}
\usepackage{titlesec}
\titleformat*{\section}{\large\bfseries}
\titleformat*{\subsection}{\normalsize\bfseries}
\titleformat*{\subsubsection}{\normalsize}

\begin{document}
\title{\LARGE \bfseries Edge states as mediators of bypass transition in boundary-layer flows}

\author[1,2]{\small{T. Khapko}}
\author[3]{T. Kreilos}
\author[1,2]{P. Schlatter}
\author[4]{Y. Duguet}
\author[5,6]{B. Eckhardt}
\author[1,2]{D. S. Henningson}

\affil[1]{Linn\'e FLOW Centre, KTH Mechanics, Royal Institute of Technology, SE-100 44 Stockholm, Sweden}
\affil[2]{Swedish e-Science Research Centre (SeRC), Sweden}
\affil[3]{Emergent Complexity in Physical Systems Laboratory (ECPS), \'Ecole Polytechnique F\'ed\'erale de Lausanne, CH-1015 Lausanne, Switzerland}
\affil[4]{LIMSI, CNRS, Universit\'e Paris-Saclay, F-91405 Orsay, France}
\affil[5]{Fachbereich Physik, Philipps-Universit\"at Marburg, D-35032 Marburg, Germany}
\affil[6]{J.M. Burgerscentrum, Delft University of Technology, NL-2628 CD Delft, The Netherlands}

\maketitle

\abstract{The concept of edge state is investigated in the asymptotic suction boundary layer in relation with the receptivity
process to noisy perturbations and
the nucleation of turbulent spots. Edge tracking is first performed numerically, without imposing any discrete symmetry, in a large computational
domain allowing for full
spatial localisation of the perturbation velocity. The edge state is a
three-dimensional localised structure recurrently characterised by a single low-speed streak that experiences erratic bursts and planar shifts. This
recurrent
 streaky structure is then compared with predecessors of individual spot nucleation events, triggered by non-localised initial noise. The present
results suggest a nonlinear
picture, rooted in dynamical systems theory, of the nucleation process of turbulent spots in boundary-layer flows, in which the localised edge state
play the role of state-space mediator. \vspace{1cm}
}

\section{Introduction}

\begin{figure}
\centering
\includegraphics[scale=0.5]{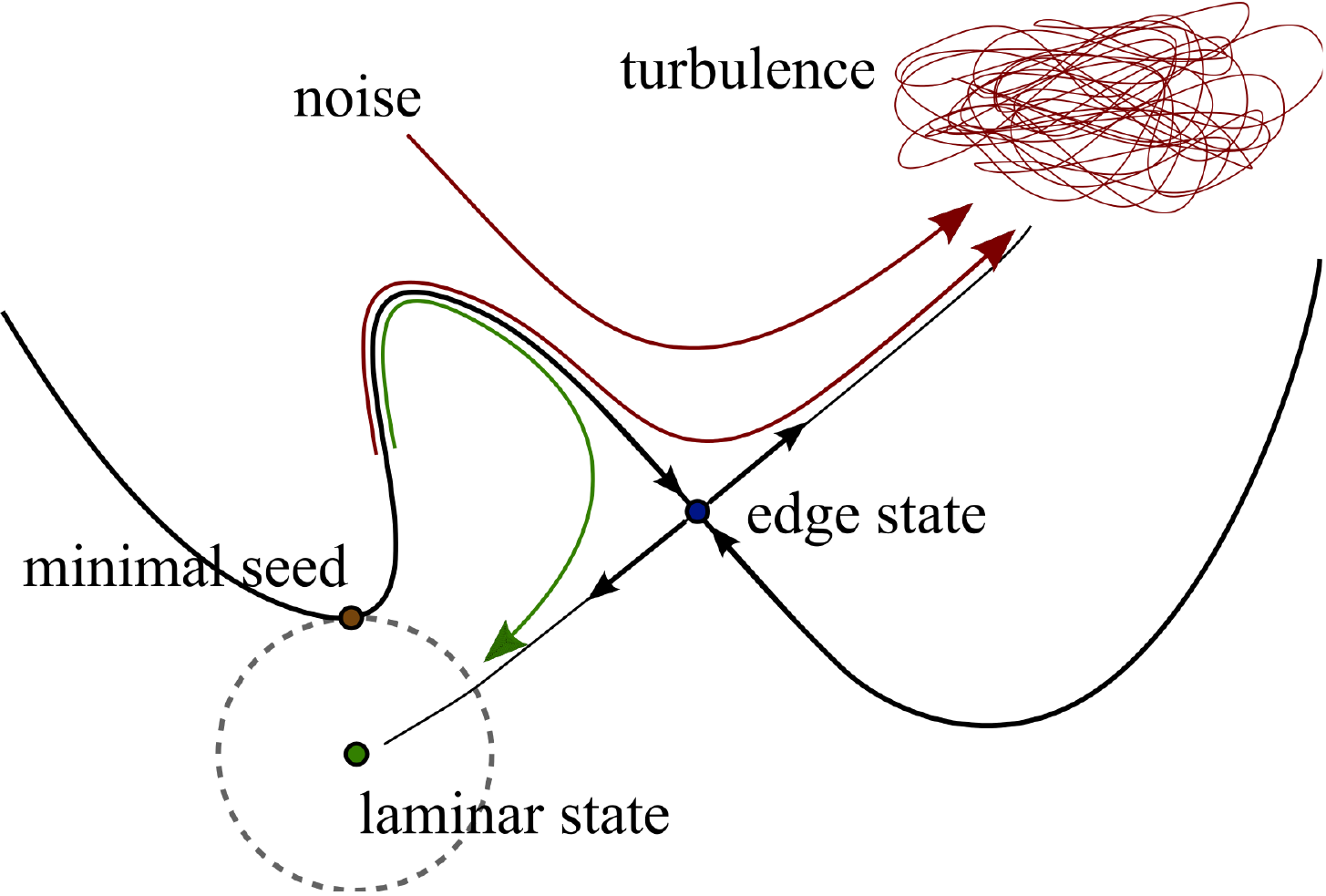}
\begin{picture}(0,0)
\put(-64,85){\boldmath$\mathcal{W}^u$}
\put(-40,25){\boldmath$\mathcal{W}^s$}
\end{picture}
\caption{\label{fig:phase-space} Sketch of the state space. The basin boundary coincides with the stable manifold $\mathcal{W}^s$ of the edge state;
$\mathcal{W}^u$ is its unstable manifold and points either towards the laminar state or to the turbulent attractor. The minimal seed is the point on
the edge closest to the laminar state in energy norm.}
\end{figure}
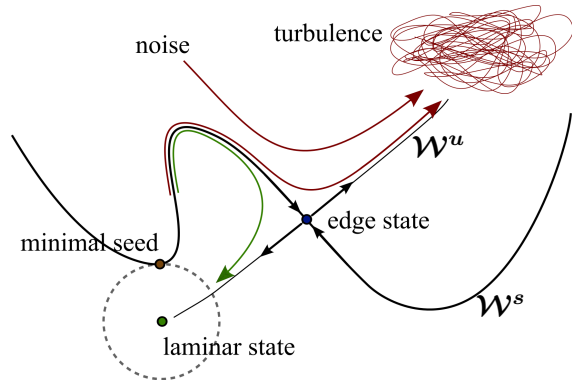

Bypass transition to turbulence in boundary-layer flows is known to proceed via the appearance of localised turbulent spots on a background
of streamwise streaks
\citep{kendall_1998,matsubara_alfredsson_2001,brandt_schlatter_henningson_2004,zaki_durbin_2005}. These spots expand aggressively and fill the plate
with turbulent motion
further downstream \citep{emmons_1951}. Bypass transition is usually described as a two-step process \citep{saric_reed_kerschen_2002}. During the
receptivity
phase,
ambient perturbations (\emph{e.g.}\ free-stream turbulence or delocalised noise) enter the boundary layer and are converted into streamwise streaks,
while in the second
phase these streaks become unstable and break down to form turbulent spots.
The receptivity phase is usually associated with linear mechanisms, while the secondary instability has traditionally been investigated by linearising around
a saturated streaky base flow \citep{andersson_brandt_bottaro_henningson_2001}. On a fundamental level, this picture of transition is not
entirely satisfying because i) it is essentially based on
disconnected linear steps, ii) streaks do not form a regular pattern and their stability must be investigated case by case, and iii) secondary
stability analysis uses spatially extended modes and therefore fails at explaining the localised nature of the ensuing spots.
Despite years of efforts to understand the nucleation
of a turbulent spot using secondary instability concepts, there has been no accepted consensus to
distinguish the perturbations leading to turbulence from those experiencing viscous decay. In wall-bounded flows such as plane Couette
or Poiseuille flows, turbulent structures can also emerge from localised disturbances, even when the laminar base flow is
linearly stable. This has lead recently to the abstract nonlinear concept of laminar--turbulent boundary in state space. According to this picture, a
 manifold of co-dimension one separates initial perturbations that rapidly decay from those
experiencing turbulent evolution \citep{itano_toh_2001, skufca_yorke_eckhardt_2006, schneider_eckhardt_yorke_2007}. This picture
carries over to extended systems, in which the edge state, \emph{i.e.}\ the relative attractor on this separating manifold,
displays robust spatial localisation
\citep{mellibovsky_meseguer_schneider_eckhardt_2009,willis_kerswell_2009,duguet_schlatter_henningson_2009,
schneider_marinc_eckhardt_2010,zammert_eckhardt_2014}. The perturbation energy of such edge states is notoriously low,
but not as low as the minimal seed considered in recent studies, itself also a point on the laminar--turbulent boundary
\citep{duguet_monokrousos_brandt_henningson_2013,kerswell_pringle_willis_2014}. Independently of its temporal dynamics
(steady, periodic or even chaotic), the edge state represents a saddle in the global state space of the system (see figure~\ref{fig:phase-space}).
Trajectories initiated arbitrarily close to its stable manifold $\mathcal{W}^s$ approach the neighbourhood of
the edge state for a short time, before being flung away along its unstable manifold $\mathcal{W}^u$. Non-localised initial perturbations are no exception
\citep{duguet_willis_kerswell_2010,schneider_marinc_eckhardt_2010}. As a consequence, even initial noise
with a carefully selected amplitude evolves into a localised seed
that later spreads like a turbulent spot. It is noteworthy that also finite-amplitude noise (with a less constraint amplitude) dissipates and 
leaves a streaky background from which isolated turbulent spots emerge (see \emph{e.g.}\ \citet{duguet_schlatter_henningson_2010} in
plane Couette flow). The visual resemblance
between localised edge states and incipient turbulent spots has also been noted, but never quantified. This leads to the following questions: can the spot nucleation process be described efficiently using this geometric state space picture involving the edge state? How far can this concept also
be extended to realistic bypass transition in boundary layer flows?

Edge states have also recently been identified in linearly stable boundary-layer flows, by considering localised initial disturbances to the laminar
base flow, and varying their amplitude, until the dynamics reaches an unstable equilibrium regime, characterised by robust spatial localisation
\citep{cherubini_depalma_robinet_bottaro_2011,duguet_schlatter_henningson_eckhardt_2012}. Spatially developing boundary layers, despite their
ubiquitous importance in nature and industry, are only
poorly adapted to edge-state computations, because the slow spatial growth complicates the search for an
asymptotically defined flow
regime. The asymptotic boundary layer (ASBL) is another boundary-layer flow, where suction at
the wall equilibrates the
growth of the boundary layer. The laminar flow solution, as well as the corresponding turbulent mean profile, are hence streamwise- and
time-independent.
The ASBL appears better suited to long-time edge state computations, because it can be simulated using periodic boundary conditions
\citep{kreilos_veble_schneider_eckhardt_2013,khapko_kreilos_schlatter_duguet_eckhardt_henningson_2013,
khapko_duguet_kreilos_schlatter_eckhardt_henningson_2013}. 

The present investigation consists of two parts. In the first part, the edge state of ASBL
is computed in a
numerical domain sufficiently large to allow for three-dimensional localisation. Its chaotic dynamics is investigated and compared with former
computations in more constrained geometries. In a second part, a set of numerical simulations of ASBL are analysed, initiated using different
realisations of random noise. The current emphasis is on identifying the signature of the edge state during spot nucleation. In an effort to match
realistic operating conditions, no special discrete symmetry has been imposed in the computations, and the amplitude of the initial noise in the
second part has been chosen arbitrarily (no bisection performed). The results
successfully confirm that localised edge states act as state space mediators during the process of spot nucleation.

\section{Flow case and numerical methodology} \label{sec:numerics}

An asymptotic suction boundary layer forms when fluid flows above a porous flat plate subject to constant uniform suction
\citep{griffith_meredith_1936}. Suction balances the natural spatial growth of the boundary layer, eventually leading to a
streamwise-independent flow. The
laminar flow solution reads $u(y)=U_\infty(1-e^{-yV_S/\nu}),~v=-V_S$, with $u$ and $v$ the streamwise and wall-normal velocity
components, $y$ the distance from the wall, $U_\infty$ the free-stream velocity and $\nu$ the kinematic viscosity of the fluid. It is realisable in
experiments \citep{antonia_fulachier_krishnamoorthy_benabid_anselmet_1988,fransson_alfredsson_2003}. The streamwise and spanwise coordinates are
denoted $x$ and $z$. The Reynolds number is defined classically as the
ratio~$Re=U_\infty/V_S$, with $V_S>0$ the suction velocity. Non-dimensionalisation is based on $U_\infty$ and on the displacement thickness
$\delta^*=\nu/V_S$. From this point onwards, all quantities are non-dimensional. The flow is linearly stable up to $Re \approx 54
370$~\citep{hocking_1975} but finite-amplitude perturbations can sustain
turbulence for $Re\ge270$~\citep{khapko_schlatter_duguet_henningson_2015}. Temporal simulations are performed using the spectral code
SIMSON~\citep{chevalier_schlatter_lundbladh_henningson_2007}. The edge state is tracked iteratively using the standard bisection procedure
of \citet{skufca_yorke_eckhardt_2006}. A new bisection is started once the separation between the two last diverging trajectories of a previous
bisection exceeds in norm a predefined threshold. Following \citet{khapko_kreilos_schlatter_duguet_eckhardt_henningson_2013,
khapko_duguet_kreilos_schlatter_eckhardt_henningson_2013}, edge computations have been performed at $Re=500$ where turbulence is an attractor. 
Two numerical domains D1 and D2 have been considered, respectively, for edge tracking and noise-induced transition: D1 has dimensions
$[L_x,L_y,L_z]=[800,15,100]$ with spectral resolution $(N_x,N_y,N_z) = (2048,121,384)$, whereas D2 has $[L_x,L_y,L_z]=[300,20,150]$ with
$(N_x,N_y,N_z) = (384,97,384)$. The large dimensions of D1 ensure full spatial localisation of the edge state while the choice for D2
results from a trade-off between capturing localisation and restricting the transition to single nucleation
events. Resolution requirements for edge tracking have already been assessed in smaller numerical domains
\citep{khapko_kreilos_schlatter_duguet_eckhardt_henningson_2013}. The slightly lower resolution used in D2 improves nucleation statistics without affecting the initial stages of transition.

\section{Localised edge state} \label{sec:invariant}

\begin{figure}
\centering
\includegraphics[width=.95\linewidth]{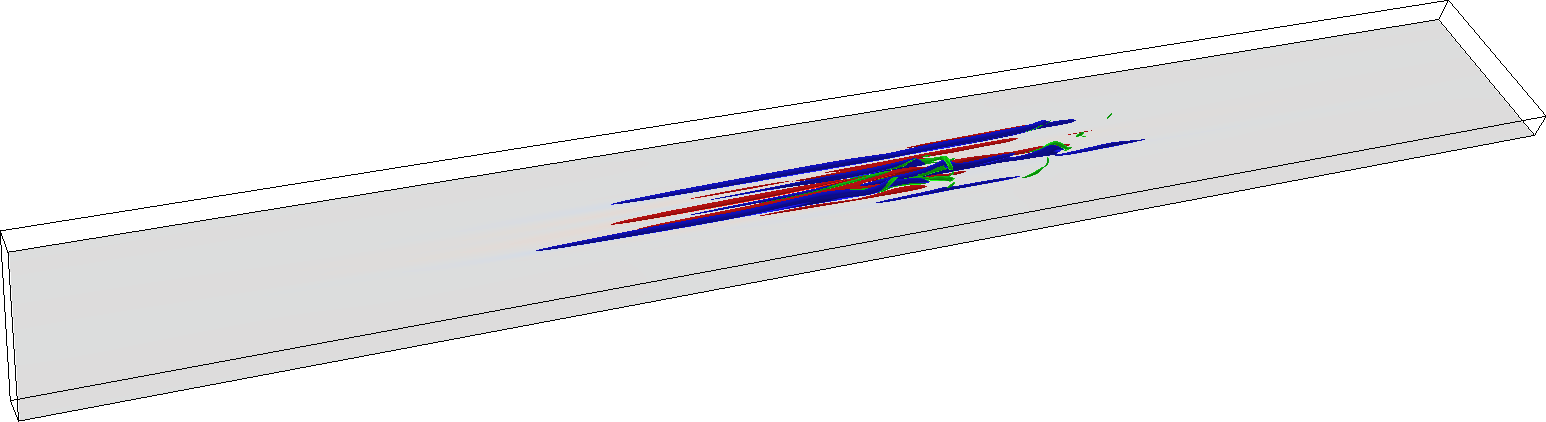}
\caption{\label{fig:edge_snapshot} Three-dimensional snapshot of the edge state ($t=2.1 \times 10^4$, see 
square in figure~\ref{fig:time_spacetime}\textit{a}). Low- (blue) and
high-speed (red) streaks shown with isosurfaces of the streamwise velocity fluctuations $u'=\pm 0.05$, and vortices with $\lambda_2=-2 \times 10^{-4}$
isosurfaces (green). Flow from left to right. The whole computational domain is shown. Supplementary movie is available in the Supplementary Material.
}
\end{figure}

\begin{figure}
\centering
\includegraphics[scale=0.33]{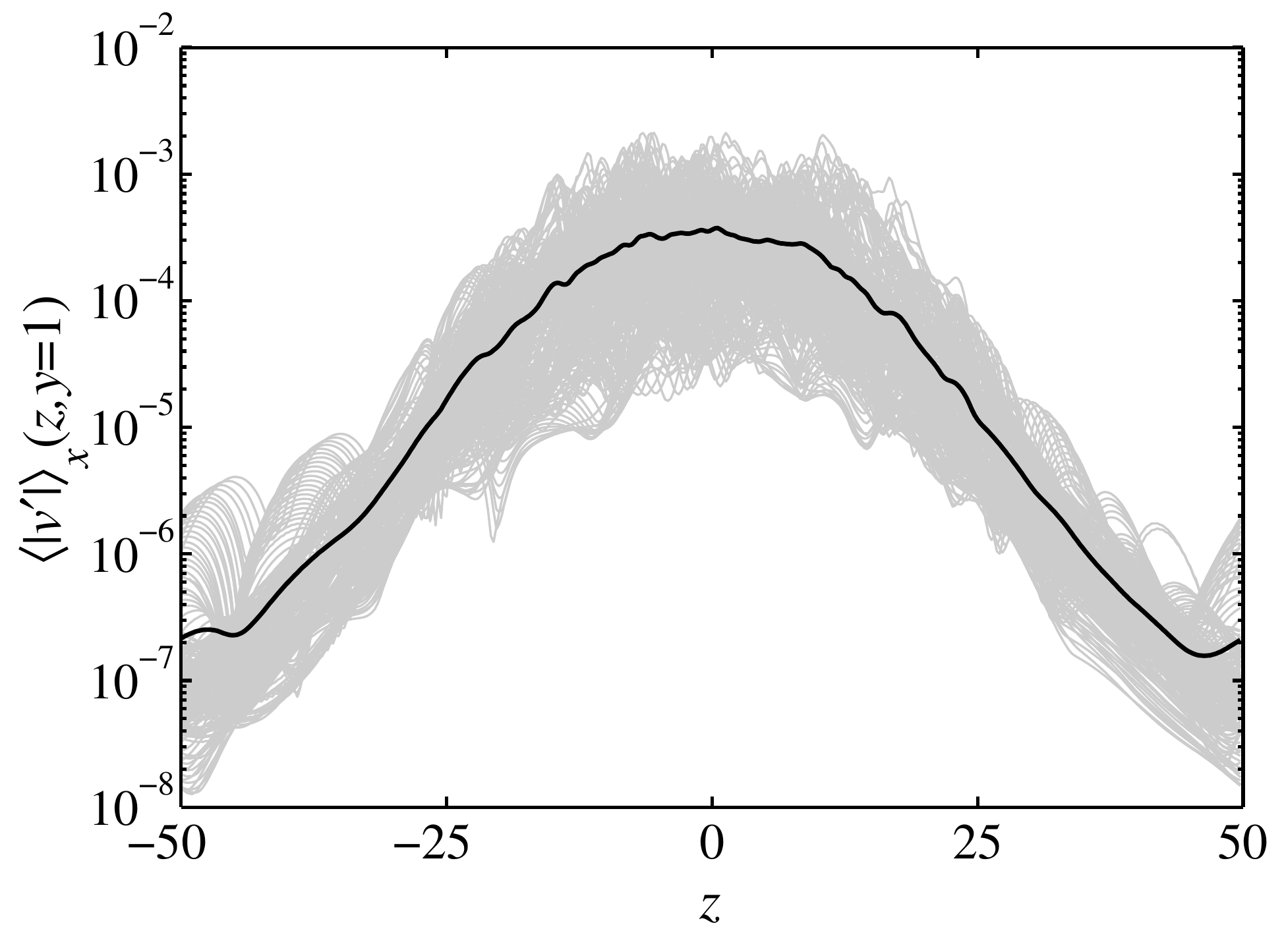}
\includegraphics[scale=0.33]{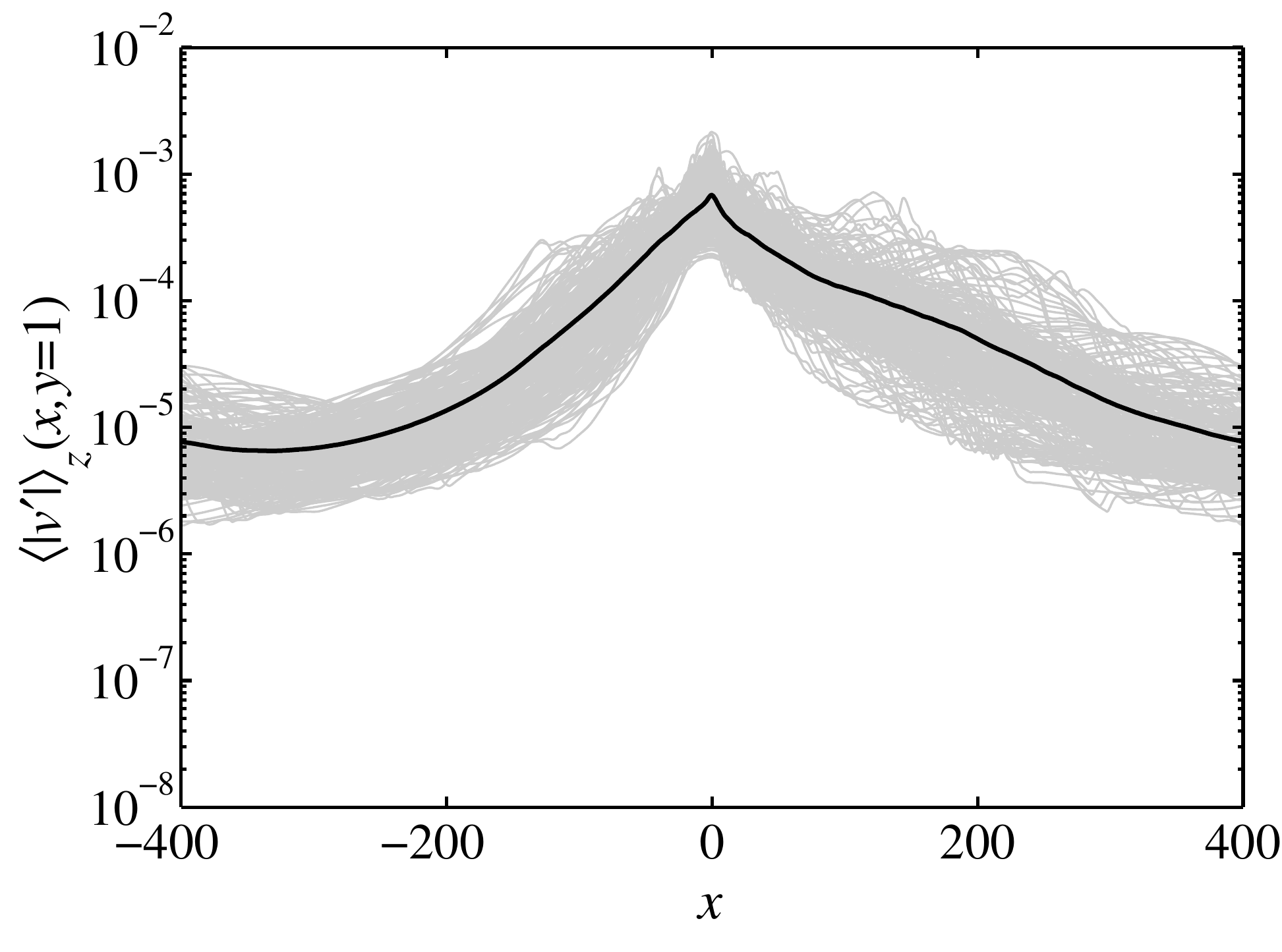}
\begin{picture}(0,0)
\put(-379,127){(\textit{a})}
\put(-191,127){(\textit{b})}
\end{picture}
\caption{\label{fig:localisation} Wall-normal fluctuations $v'(y=1)$ averaged in the streamwise (a) or spanwise direction (b).
Instantaneous (grey) \emph{vs.}\ time-averaged (black) profiles.}
\end{figure}
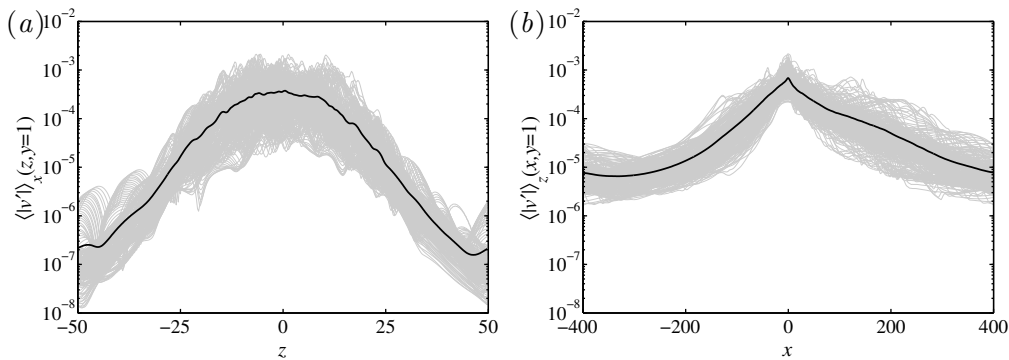

\begin{figure}
\centering
\includegraphics[scale=0.33]{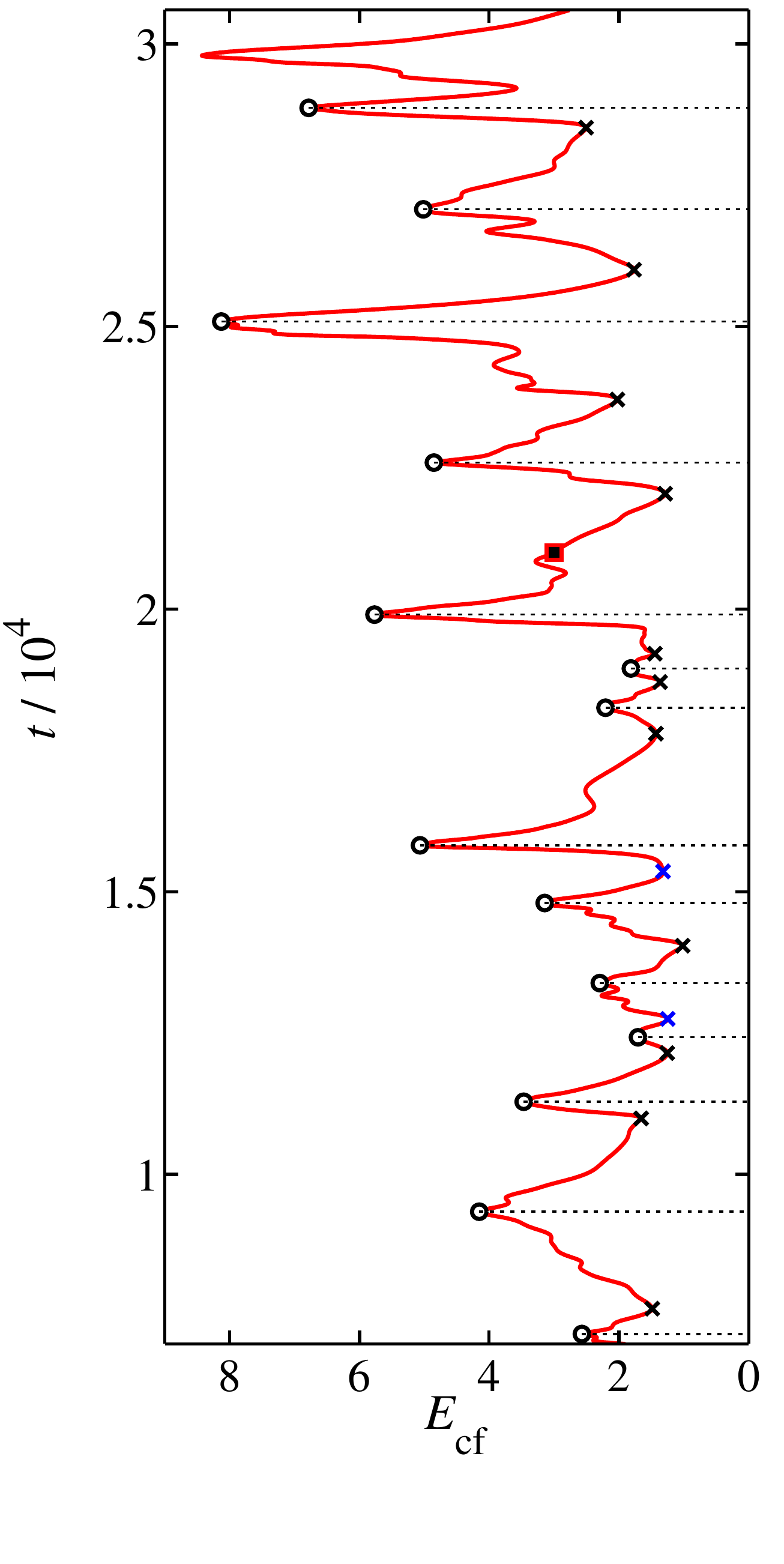}
\includegraphics[scale=0.33]{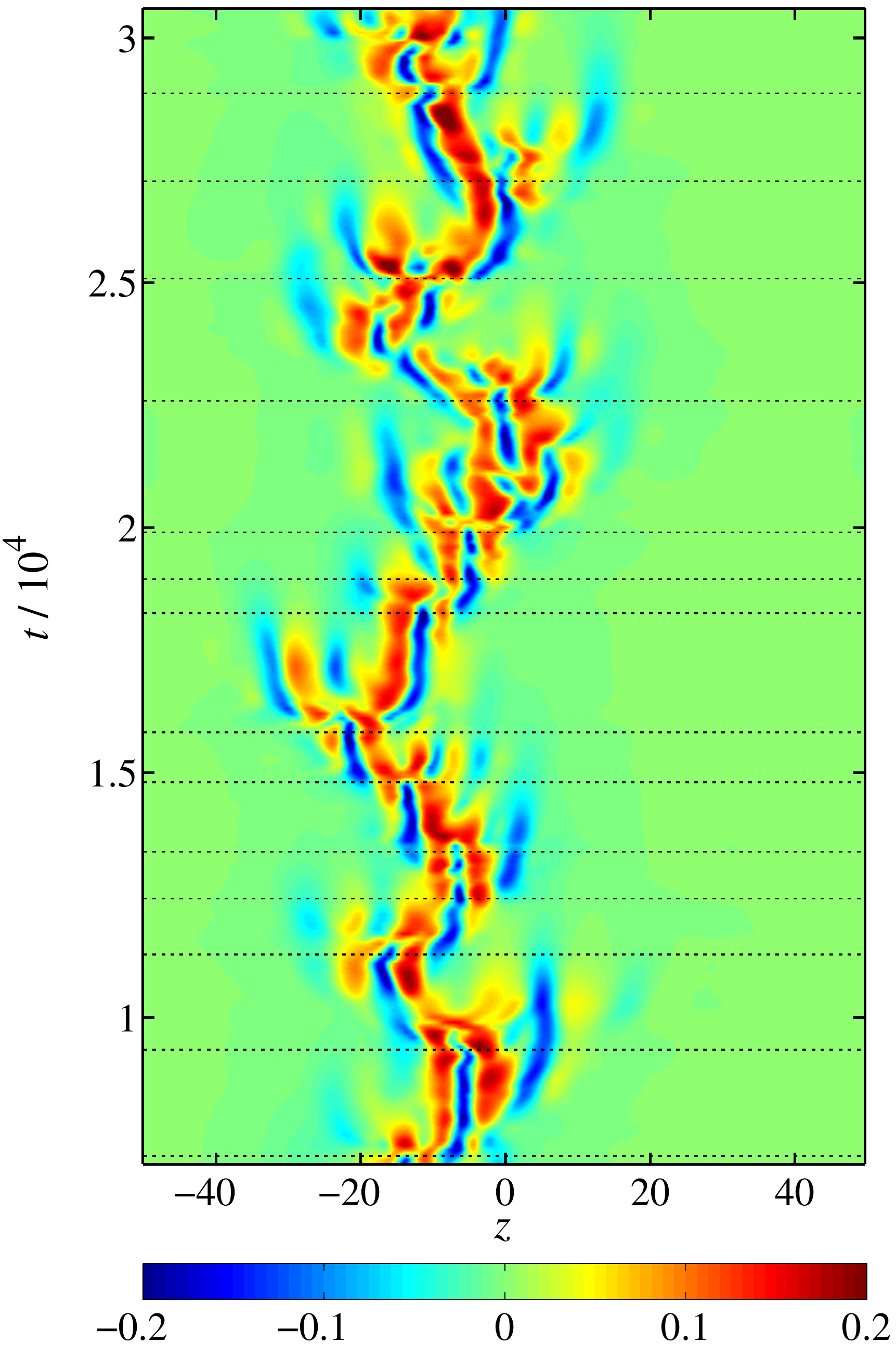}
\begin{picture}(0,0)
\put(-290,239){(\textit{a})}
\put(-162,239){(\textit{b})}
\end{picture}
\caption{\label{fig:time_spacetime} (\textit{a}) Cross-flow energy $E_\mathrm{cf}(t)$. Open
circles (resp. crosses) correspond to maxima (resp. minima) of $E_\mathrm{cf}$. The snapshot in
figure~\ref{fig:edge_snapshot} is marked with a square, those from figure~\ref{fig:edge_vs_nucleation} with blue crosses. (\textit{b})
Space--time $(z,t)$ diagram
of $\langle u' \rangle_{x}(y=1)$. Averaging in $x$ is performed here locally in a neighbourhood around the active core of the state.
}
\end{figure}
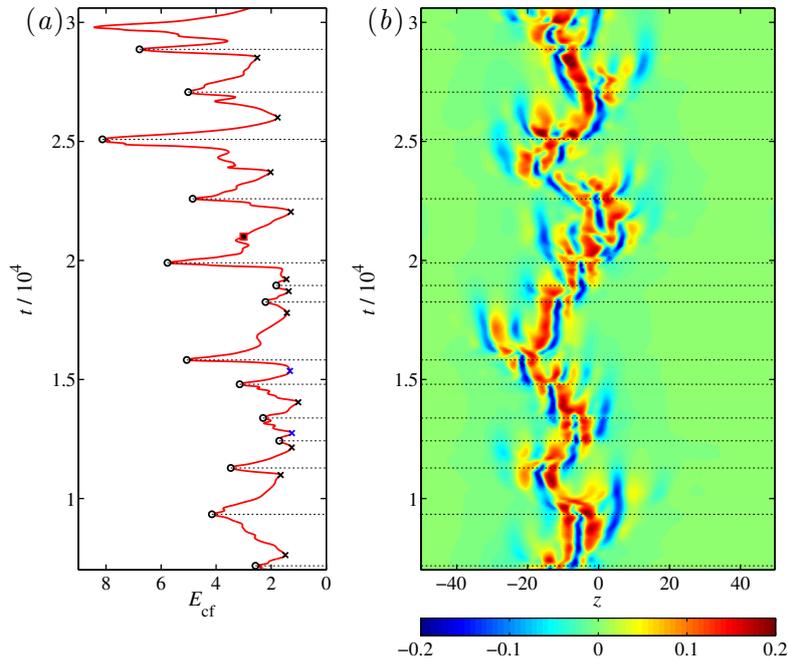

Edge tracking results asymptotically in a fully localised structure. A three-dimensional snapshot of the edge state is shown
in figure~\ref{fig:edge_snapshot}. It consists of a few low- and high-speed streaks and streamwise vortices
located mostly along one main low-speed streak. The structure resembles a turbulent spot in its early development stage, except that it neither decays
nor grows in size. For slightly higher initial amplitudes, the same flow structure quickly breaks down into turbulence that spreads throughout the 
whole domain. Importantly, this edge state is localised in the three spatial dimensions. In the wall-normal direction it is
contained within the boundary layer. Full localisation in the
wall-parallel directions is demonstrated in figure~\ref{fig:localisation} by considering the norm of the wall-normal fluctuations~$|v'|$ at $y=1$.
The averages of $|v'|$ in the $x$ or $z$ direction, denoted with $\langle \cdot \rangle$, are then computed to assess the spanwise and
streamwise localisation (see figure~\ref{fig:localisation}).
The drop
in magnitude in $z$ by four decades is consistent with an exponential decay, whereas the decay by two decades in $x$ is
compatible both with
exponential or power-law decay \citep{zammert_eckhardt_2014}. The temporal dynamics of the edge state is monitored in
figure~\ref{fig:time_spacetime}(\textit{a}) using the cross-flow energy
\begin{equation}
E_{\mathrm{cf}}(t)= \int_{\Omega} (v'^2 + w'^2)\, \mathrm dx\, \mathrm dy\, \mathrm dz \equiv E_v+E_w \ ,
\end{equation}
based on the wall-normal and spanwise velocity fluctuations. This quantity leaves aside the fluctuations associated with
the
streamwise streaks, characterised by weak decay rates. Variations of $E_\mathrm{cf}$ remain chaotic but display a clear alternation of calm phases
with bursting phases.
The spatio-temporal dynamics of the streaks is relatively simple once displayed in a space--time diagram as in
figure~\ref{fig:time_spacetime}(\textit{b}). 
The quantity visualised here is the streamwise velocity fluctuation $u'(y=1)$ averaged locally over $50$ streamwise units. When $E_\mathrm{cf}$ is low, only one low-speed streak
remains active, which defines the recurrent `active core' of the edge state. The active low-speed streak is flanked by staggered streamwise
vortices. A
sinuous mode develops and ultimately breaks down the streak (see left column in figure~\ref{fig:edge_vs_nucleation}), in association with a burst
of $E_\mathrm{cf}$ in figure~\ref{fig:time_spacetime}(\textit{a}).
In the breakdown process new vortices are created around the active region, regenerating new streaks. One of them takes the role of the
active core and the cycle is closed modulo a spanwise shift. This is confirmed in figure~\ref{fig:time_spacetime}(\textit{b}) which shows erratic
shifts in $z$ over periods of $\approx 10^3-10^4$ time units. The whole regeneration cycle is the localised counterpart of the dynamics described in
shorter numerical domains of ASBL \citep{kreilos_veble_schneider_eckhardt_2013,khapko_kreilos_schlatter_duguet_eckhardt_henningson_2013,
khapko_duguet_kreilos_schlatter_eckhardt_henningson_2013}, and bears many similarities with the self-sustaining cycle of near-wall turbulence
\citep{hamilton_kim_waleffe_1995}. The fluctuations reaching higher up in the
boundary layer are shed downstream from
the active
core, where the shear is not strong enough to sustain them. This mechanism, also reported in spatially developing boundary layers
\citep{duguet_schlatter_henningson_eckhardt_2012}, maintains the $x$-localisation of the edge state.

\section{Nucleation process} \label{sec:nucleations}

Sinuous modes growing on low-speed streaks have often been reported in experiments and computations of bypass
transition \citep{brandt_schlatter_henningson_2004,yoshioka_fransson_alfredsson_2004}. Varicose modes of instability have also been predicted and
observed in boundary layer flows but are outweighed statistically by the sinuous ones \citep{schlatter_brandt_delange_henningson_2008}. The visual
similarities between the edge-state dynamics and pre-nucleation episodes in bypass transition suggest that the emergence of streaks may be regarded as
the approach
to $\mathcal{W}^s$, the stable manifold of the edge state. A crucial difference lies in the treatment of the sinuous modes, traditionally interpreted as a growing departure from a streaky non-sinuous base 
flow. The current sinuous oscillations are inherently part of the dynamics of the edge state. Constraining the dynamics to remain on $\mathcal{W}^s$
on the one hand leads
to break-up and later reformation of exactly one active streak. Unconstrained computations on the other hand do not lead to one active streak
reforming after breakdown: either the streak does not reform and the flow relaminarises, or several low-speed streaks emerge which cascade into
turbulent motion and finally lead to an aggressively expanding turbulent spot. Both scenarios take the flow away from the edge
state,
which is consistent with a state space escape along either direction of $\mathcal{W}^u$.
\begin{figure}
\centering
\frame{\includegraphics[scale=0.33]{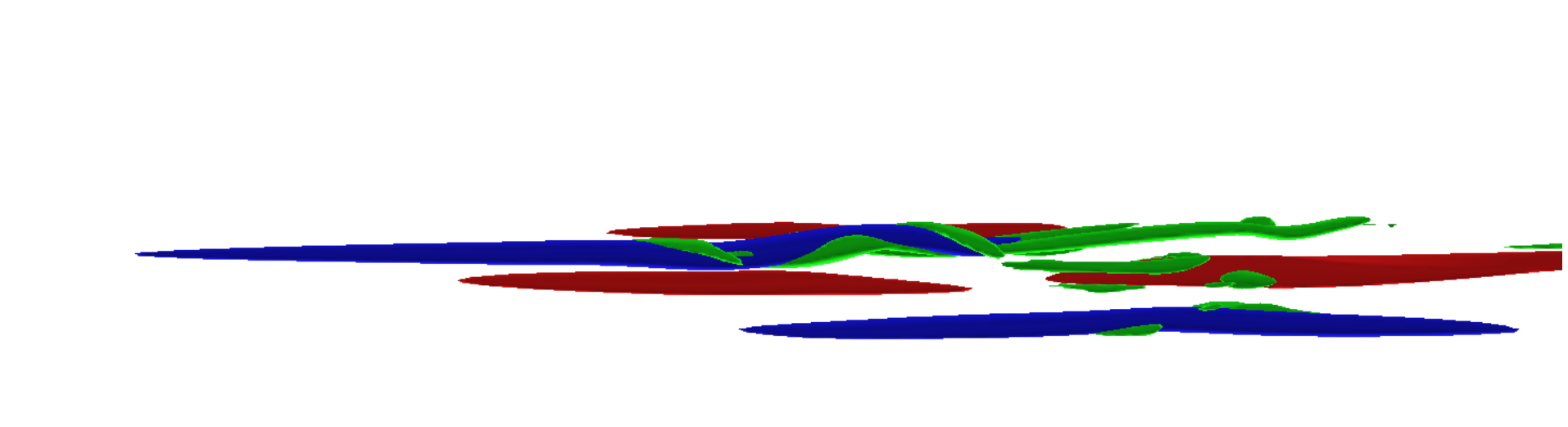}} \hspace{.1cm}
\frame{\includegraphics[scale=0.33]{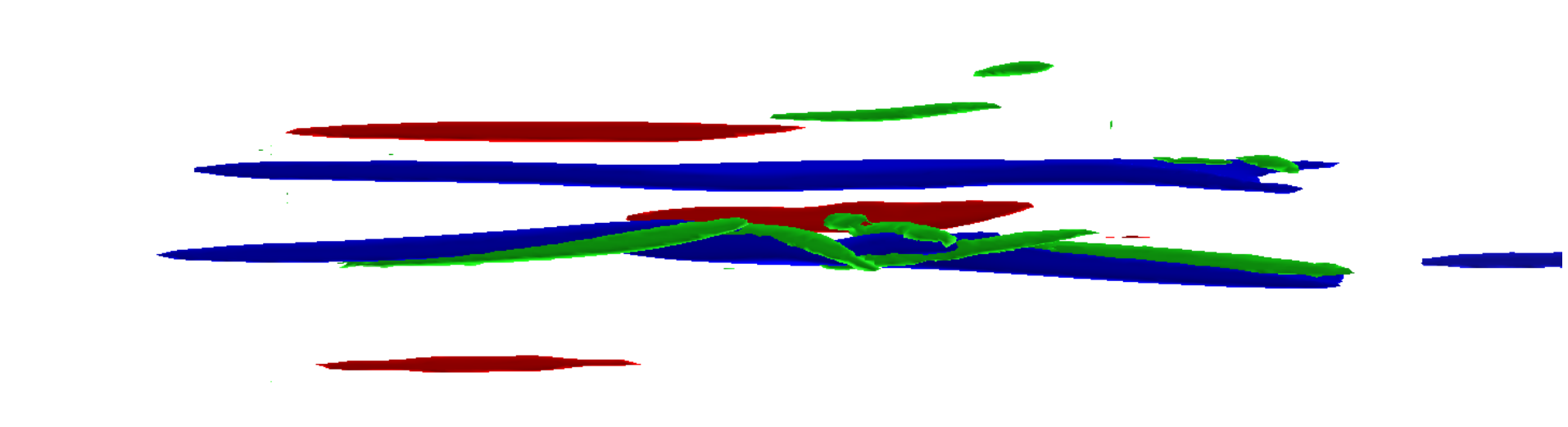}}\\
\frame{\includegraphics[scale=0.33]{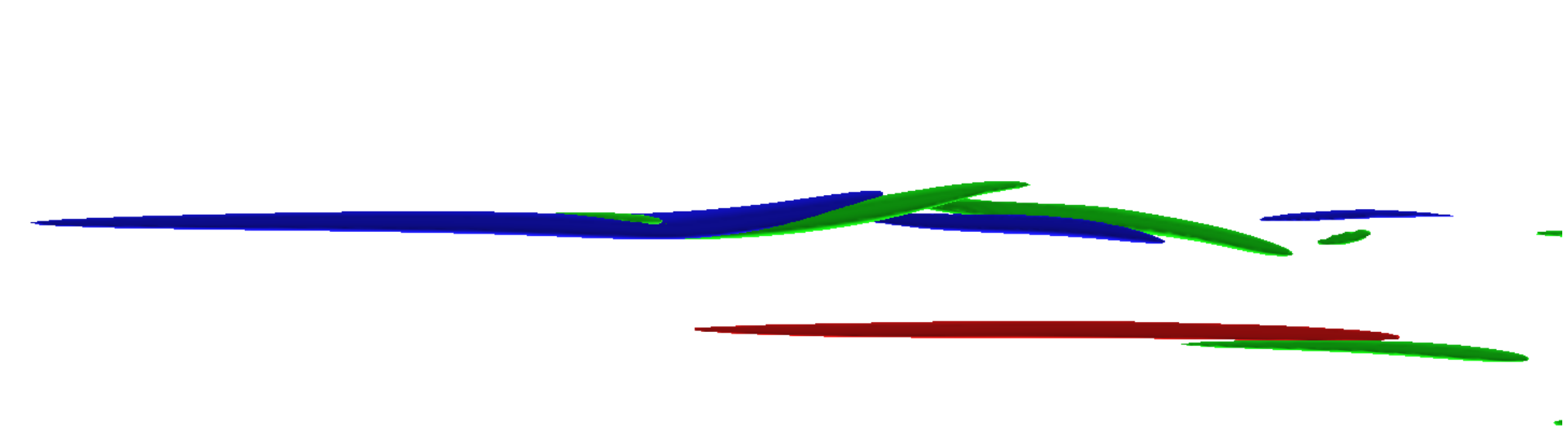}} \hspace{.1cm}
\frame{\includegraphics[scale=0.33]{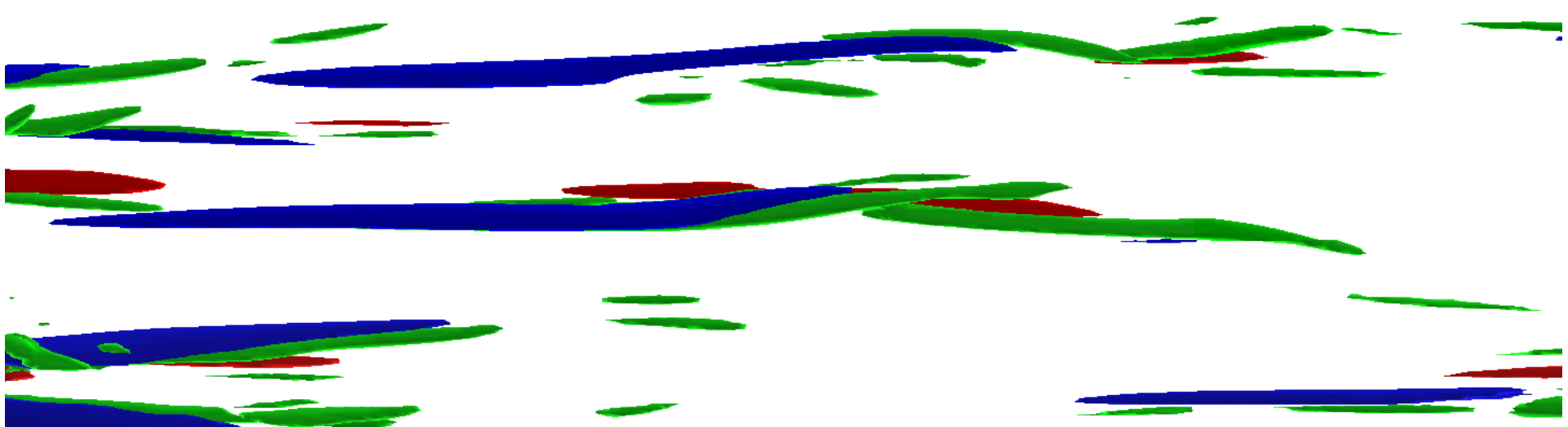}}
\caption{\label{fig:edge_vs_nucleation} Comparison of the edge state during calm phases ($t=12700,~15400$) and pre-nucleation events in noise-induced
transition. Visualisation similar to figure~\ref{fig:edge_snapshot}, but with different isovalues ($u'=\pm
0.13$ and $\lambda_2=-4\times10^{-4}$). Only parts of
the computational domains D1 and D2 (approximately $140 \times 40$) are shown. Flow from left to right.}
\end{figure}
In order to test this hypothesis, we consider a series of simulations of transition ASBL at $Re=500$ with non-localised noise as an initial condition,
and investigate whether the edge state can be, even transiently, identified in these simulations. The noise consists of a random
superposition of Stokes modes spanning the whole domain with length scale on the order of $10$ \citep{chevalier_schlatter_lundbladh_henningson_2007}.
The magnitude for the noise amplitude is
chosen
to ensure transition through nucleation and
growth of turbulent spots, but
it is not optimised to result in trajectories that spend a long time near the
edge (as could be done using edge tracking) since we did not want to artificially
force visits to the edge states. As noted in section~\ref{sec:numerics}, the numerical domain D2 is chosen
sufficiently large for localised spots to emerge, but still small enough so that typically only one or two distinct localised spot events are
observed.

\begin{table}
\centering
  \begin{tabular}{l|c|c}
    Characteristic         & edge state in calm phase & pre-nucleation events \\ \hline \hline
    streak width $[\delta^*]$           & $3.46 \pm 0.52$   & $3.75 \pm 0.63$  \\ \hline
    streak intensity $[U_\infty]$        & $0.25 \pm 0.03$  & $-0.25 \pm 0.04$ \\ \hline
    instability wavelength $[\delta^*]$ & $33 \pm 10$       & $27.5  \pm 6$  \\ \hline
  \end{tabular}
\caption{Quantitative comparison of edge state and noise-induced pre-nucleation events.} \label{tab:characteristics}
\end{table}

\begin{figure}
\centering
\includegraphics[scale=0.33]{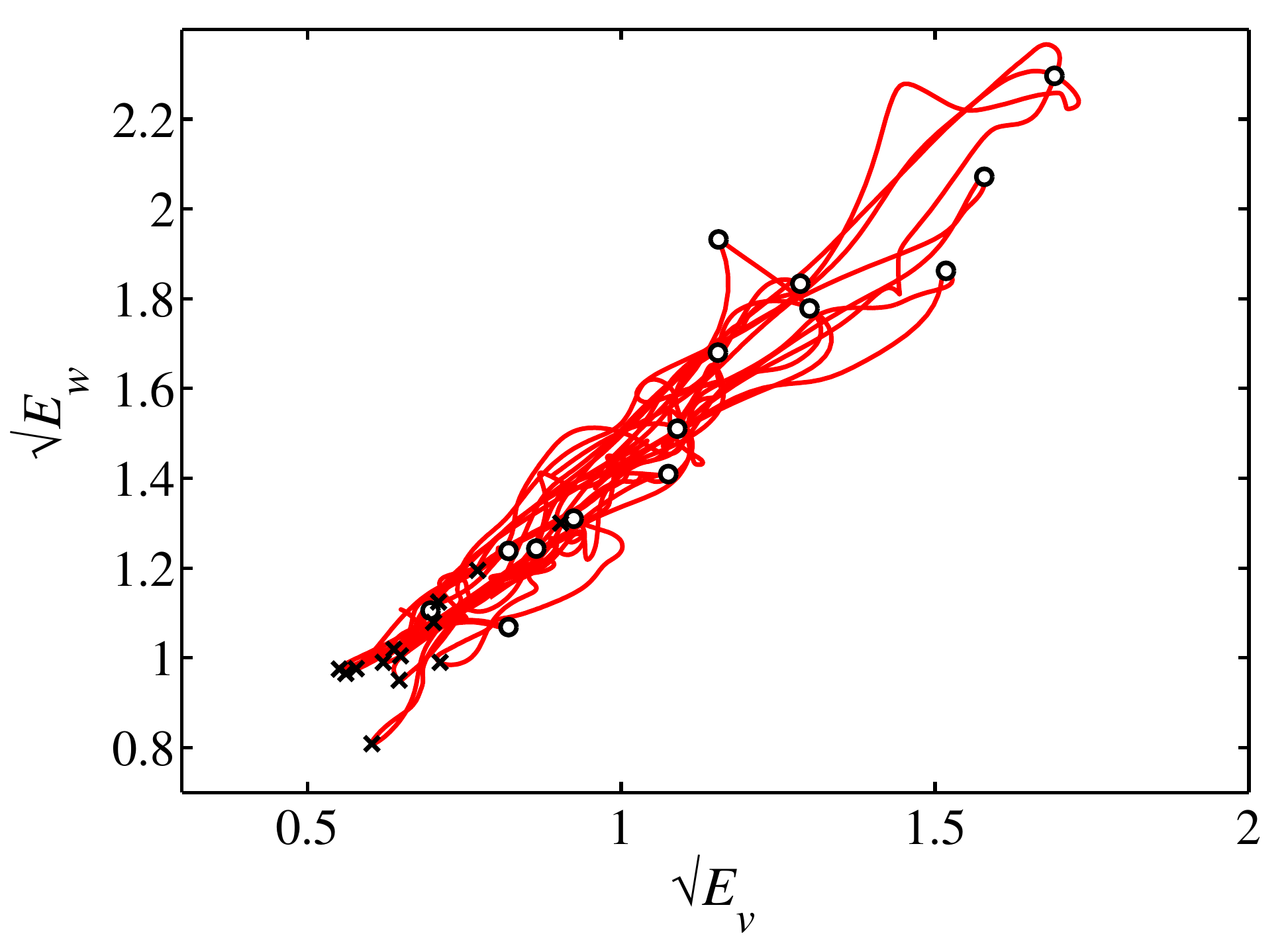}
\includegraphics[scale=0.33]{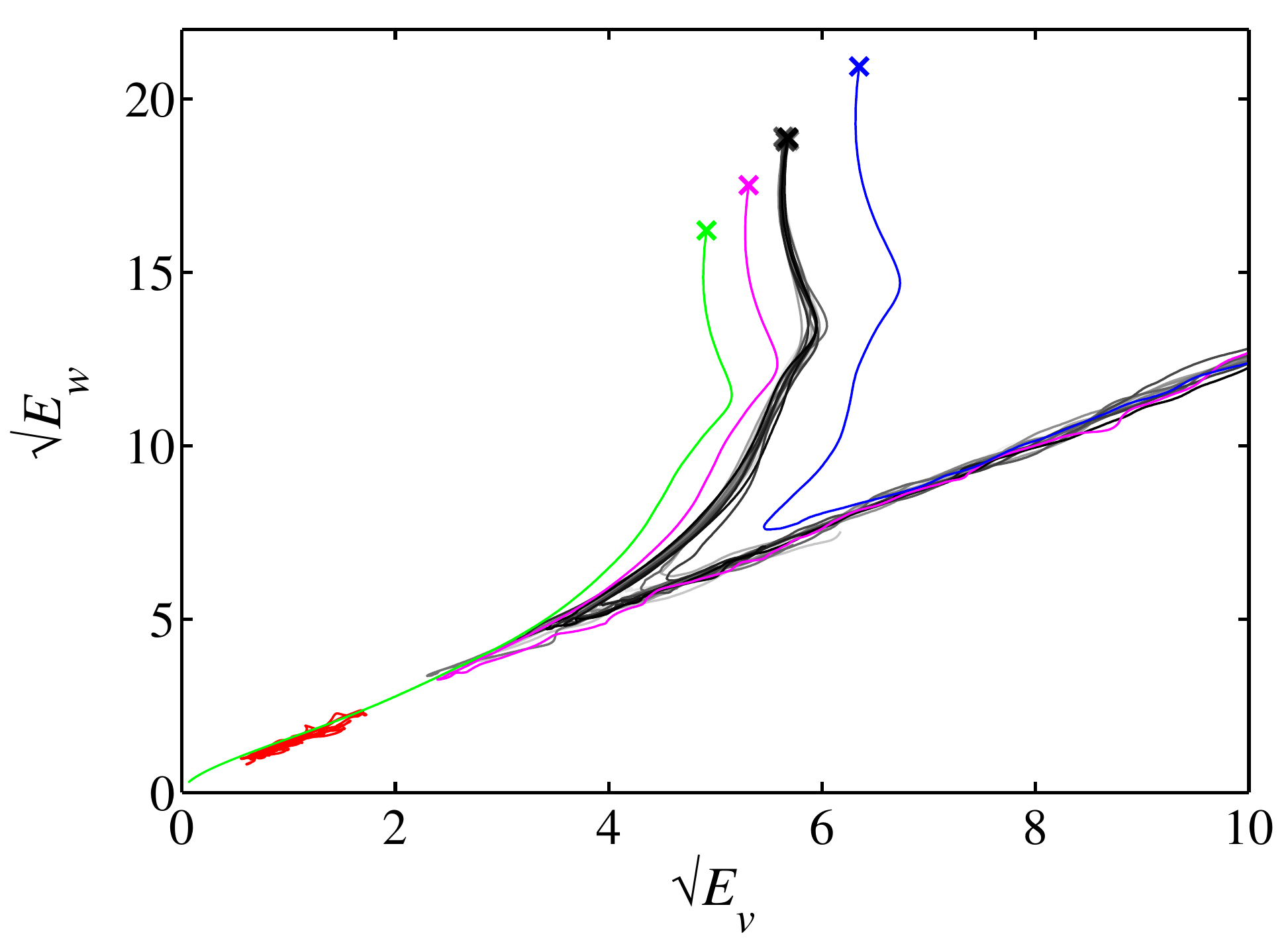}
\begin{picture}(0,0)
\put(-376,120){(\textit{a})}
\put(-187,120){(\textit{b})}
\end{picture}
\caption{\label{fig:phase-space_ASBL} State space projection on $\sqrt{E_v}$ and $\sqrt{E_w}$.
(\textit{a}) Projection of the chaotic edge
state. Circles
and crosses indicate extrema of $E_\mathrm{cf}$ as in figure~\ref{fig:time_spacetime}(\textit{a}). (\textit{b}) Transition from four different noise
amplitudes (green, magenta, shades
of grey and blue colours in the plot), with most of the 
trajectories corresponding to one of the noise levels (grey).
Noisy initial conditions with crosses, edge state in red. Supplementary movie is available in the Supplementary Material.
}
\end{figure}
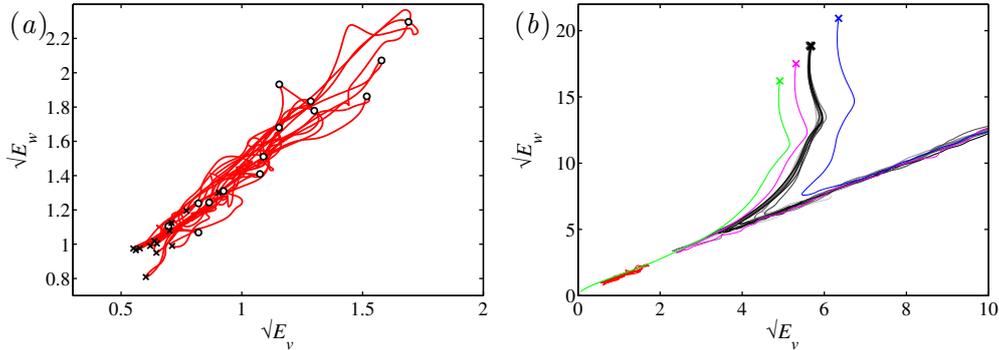

We begin with a qualitative comparison in figure~\ref{fig:edge_vs_nucleation} between well-chosen snapshots on the edge trajectory and local
precursors of turbulent spots. 
Both instances involve one low-speed streak with sinuous instabilities with similar size and magnitude. More quantitative evidence that the edge state
is actually embedded in the noise simulations is based on three-dimensional measurements (statistics are based on $13$ instances corresponding to all local minima of $E_\mathrm{cf}$). We report in the two set-ups the spanwise width of the most
intense low-speed streak, its intensity (deviation from the laminar flow) and the streamwise wavelength of the growing sinuous instability. The
quantitative similarity between the
statistical averages reported in table~\ref{tab:characteristics} is remarkable.
Further quantitative evidence that noise-initiated trajectories do visit the neighbourhood of the edge state comes from energetic projections of the
state space. In figure~\ref{fig:phase-space_ASBL}, the dynamics is projected on the
$(\sqrt{E_v},\sqrt{E_w})$ plane. In this projection, the chaotic edge state is a convoluted object residing in a small bounded region of the state
space (see
figure~\ref{fig:phase-space_ASBL}\textit{a}).
Trajectories with the lowest levels of initial noise (green line in figure~\ref{fig:phase-space_ASBL}\textit{b}) continue
towards the laminar state while the flow slowly laminarises, suggesting that the corresponding initial conditions are within the basin of attraction of the laminar state. For most initial conditions with higher amplitude, apparent approaches to the edge state in figure~\ref{fig:phase-space_ASBL}(\textit{b}) are
followed by an abrupt increase in energy: the corresponding initial conditions belong to the basin of attraction of the turbulent state. The cloud of
initial conditions is hence staggered along $\mathcal{W}^s$. Several simultaneous nucleations result in higher energies and, thus, in a larger
distance to the edge in figure~\ref{fig:phase-space_ASBL}(\textit{b}). Therefore, the distance of the closest approach is
smallest for the
trajectories involving
one single nucleation event. Once the spots appear, they grow in size, invading the domain,
with the energy steadily growing in figure~\ref{fig:phase-space_ASBL}(\textit{b}). In the
current finite-size system the energy saturates when the entire volume has become
turbulent. Since
the latter is domain-dependent, we choose not to show it in figure~\ref{fig:phase-space_ASBL}(\textit{b}). The geometric structure emerging from
figure~\ref{fig:phase-space_ASBL}(\textit{b}) is then directly comparable with figure~\ref{fig:phase-space}. All trajectories in
figure~\ref{fig:phase-space_ASBL}(\textit{b}), despite starting from
different
parts of the state space (the variability of $E_\mathrm{cf}(t=0)$ is here approximately $70 \%$), approach the location of the edge state and
leave
together along a single direction. This is consistent with an approach
along $\mathcal{W}^s$ (characterised by a very large effective dimension) and an escape along the one-dimensional $\mathcal{W}^u$.

\begin{figure}
\centering
\includegraphics[scale=0.33]{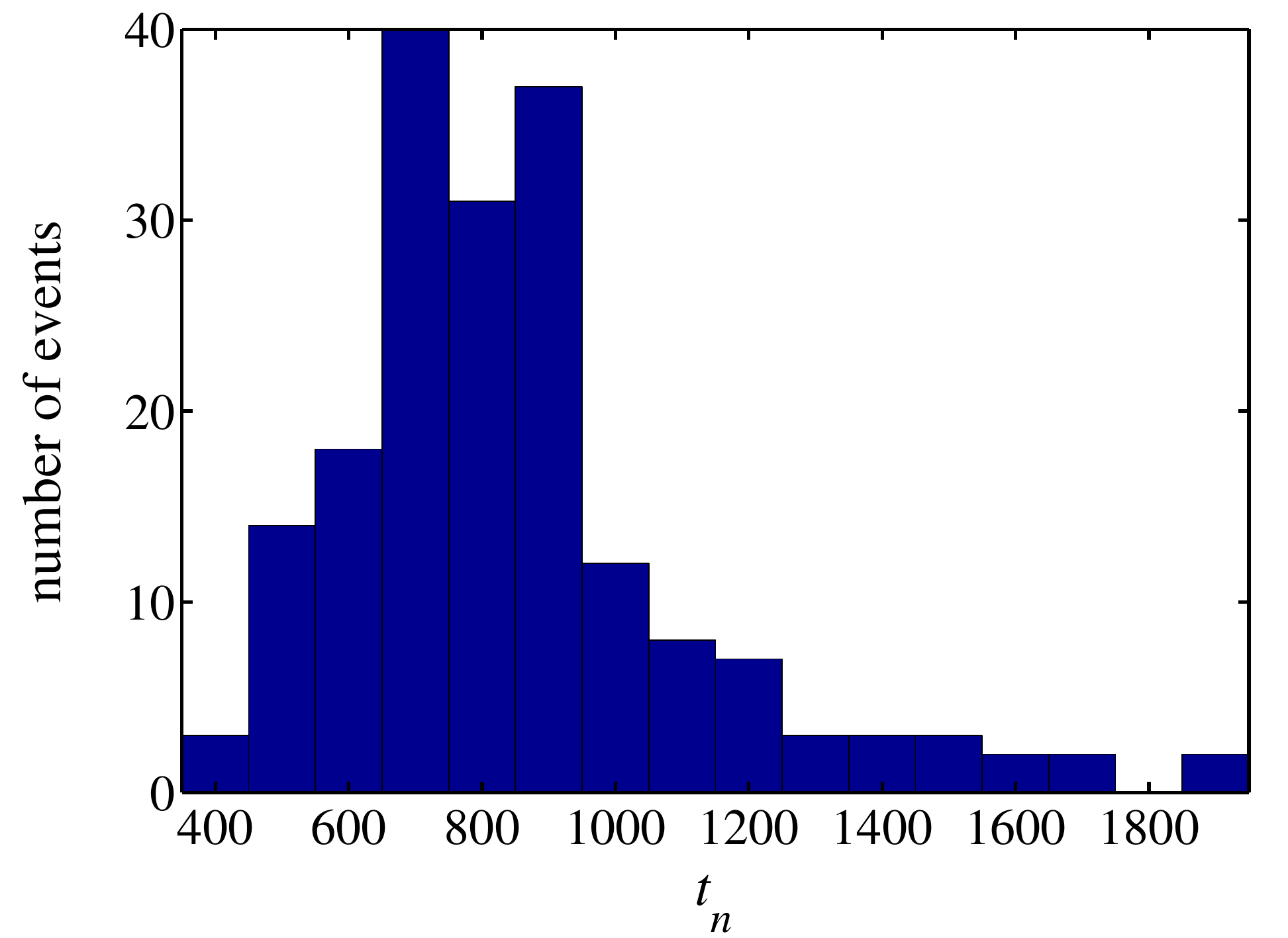}
\includegraphics[scale=0.33]{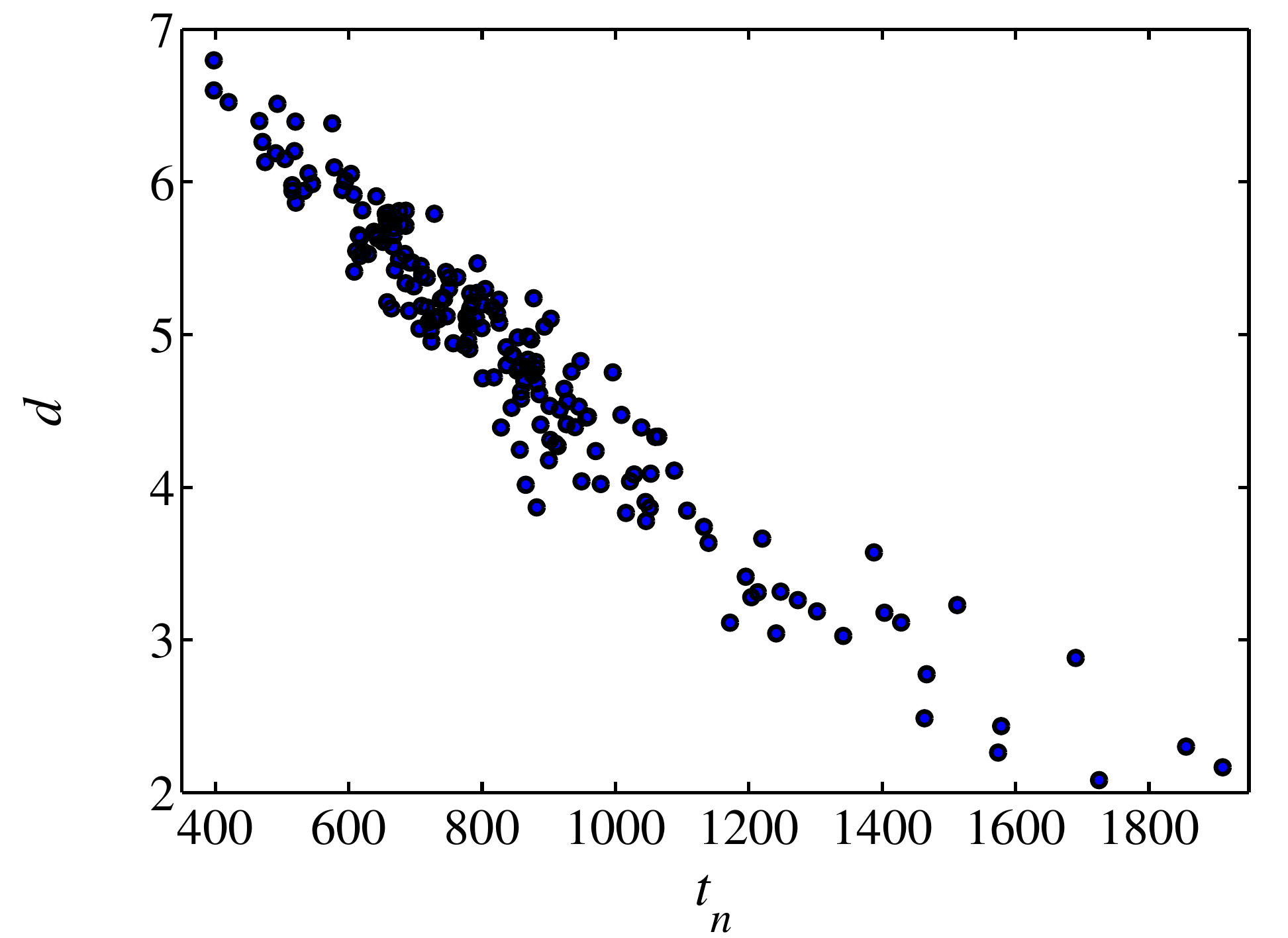}
\begin{picture}(0,0)
\put(-376,120){(\textit{a})}
\put(-192,120){(\textit{b})}
\end{picture}
\caption{\label{fig:nucleation_time} (\textit{a}) Distribution of the nucleation events in time obtained for the same initial noise level.
(\textit{b}) Nucleation time $t_n$ \emph{vs.}\ the distance of the closest approach to the edge $d$.}
\end{figure}
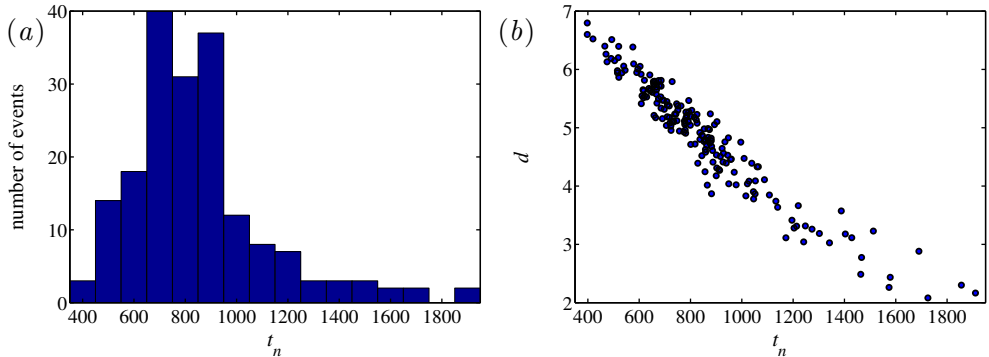

The time it takes to nucleate a turbulent spot starting from noise is different for each individual state space trajectory. This \emph{nucleation
time}
$t_n$ is defined as the time at which a predefined threshold in $E_\mathrm{cf}=70$ is reached. A histogram of $t_n$ for different trajectories
starting from the same noise level is shown in
figure~\ref{fig:nucleation_time}(\textit{a}). Its exact shape depends on the statistical distribution of initial conditions
and details of the dynamics. 
However, we infer from the data in figure~\ref{fig:phase-space_ASBL}(\textit{b}) that closer approaches to the edge state lead to delayed transition. This is verified in
figure~\ref{fig:nucleation_time}(\textit{b}) by plotting for each noise realisation the measured nucleation time versus the associated minimal
distance to the calm phase on the edge in the $(\sqrt{E_v},\sqrt{E_w})$ plane. The trend is clear
and, despite the nonlinear
context, fully consistent
with the linear approach to a saddle point: the
closer a trajectory comes to the edge state, the longer the time it spends in its vicinity, hence the later it reaches the turbulent attractor. This
behaviour, expected in the direct (linearised) neighbourhood of a saddle point, is confirmed here at a finite distance from the saddle. Note that in
an
experimental set-up, a distribution of nucleation times translates into a distribution of nucleation positions because of the mean advection. 

\section{Discussion and conclusions} \label{sec:conclusions}

Computation of the edge state in ASBL in an extended numerical domain has lead to a flow state localised in the three spatial directions. Its
dynamics is chaotic in time and involves recurrent visits to an active core consisting of one low-speed streak developing sinuous instabilities. The regeneration mechanism identified is strongly reminiscent of similar
computations in constrained
numerical domains \citep{kreilos_veble_schneider_eckhardt_2013,khapko_kreilos_schlatter_duguet_eckhardt_henningson_2013}, and appears thus generic for ASBL. Besides, localised recurrent low-speed streaks subject to
sinuous instabilities have also been reported as edge states in spatially developing boundary layer~\citep{duguet_schlatter_henningson_eckhardt_2012}
with imposed spanwise symmetry, suggesting possible direct extensions to a wider class of boundary layer flows.

The classical description of a nucleation event leading to a turbulent spot is based on the linear instability of some well-chosen streak state
selected by the receptivity process. This can here be reformulated within a more global state space picture where the central mediator is the edge
state along with his stable and unstable manifolds $\mathcal{W}^s$ and $\mathcal{W}^u$. The approach along $\mathcal{W}^s$ incorporates both the
classical receptivity stage and also the emerging localisation, while the escape along $\mathcal{W}^u$ captures all the mechanisms, formerly described
by secondary instability concepts, now in a localised context. The concept of escape from the edge was used recently to model the nucleation process
\citep{kreilos_khapko_schlatter_duguet_henningson_eckhardt_2015}, without taking into account the receptivity phase encoded in the approach to
$\mathcal{W}^s$.\\

The present study demonstrates that this picture does not only hold for very carefully selected initial conditions designed to lie on $\mathcal{W}^s$,
but that it is robust enough for a much wider class of perturbations. Careful unbiased analysis of a set of simulations initiated by
non-localised noise
convincingly demonstrates that a robust coherent structure, quantitatively analogous to the recurrent streak state on the edge, is also closely
visited after the initial phase where the noise dissipates. Former evidence for the presence of a well-defined finite-amplitude streaky state as
precursor to spot nucleation appeared in various shear flows studies, for trajectories starting either from a minimal seed
\citep{duguet_monokrousos_brandt_henningson_2013} or from free-stream turbulence \citep{schlatter_brandt_delange_henningson_2008}. A finer
representation of the state space geometry with a proper metrics, taking into account the chaotic behaviour on the edge, its expanding Lyapunov
directions and the translational 
invariance of the system, would improve our representation of the system and the measure of how close the edge state is approached. The minimal seed
is another well-defined point in this geometry, also connected to the edge state, and its dynamic role in a realistic transitional flow also deserves
further study. This view on the bypass-transition process appears now generic to most wall-bounded shear flows, where transition can be considered as
an initial value problem in a temporal set-up. The main challenge is to reconcile such a set-up with permanent excitations
fluctuating both in time
and space. The extension of an abstract state-space description to spatially developing boundary layer flows still faces further open questions. While
the concept of localised edge state is useful to describe the formation of one individual turbulent spot, the case of truly extended systems
(\emph{i.e.}\ larger than the typical correlation distance) demands a paradigm shift: as multiple simultaneous nucleation events occur at 
remote (ideally uncorrelated) locations on the plate, the description of the system in terms of a unique dynamical system reaches its limits. The price to pay is to trade the deterministic picture for a statistical description of the nucleation process involving the definition of a probabilistic nucleation rate. \\

We acknowledge that the results of this research have been achieved using the PRACE-3IP project (FP7 RI-312763) resource ARCHER based in UK at
EPCC. Some additional runs were performed using the resources provided by SNIC (Swedish National Infrastructure for Computing).


\begin{thebibliography}{34}
\expandafter\ifx\csname natexlab\endcsname\relax\def\natexlab#1{#1}\fi

\bibitem[Andersson {\em et~al.\/}(2001)Andersson, Brandt, Bottaro \&
  Henningson]{andersson_brandt_bottaro_henningson_2001}
{\sc Andersson, P., Brandt, L., Bottaro, A. \& Henningson, D.~S.} 2001 {On the
  breakdown of boundary layer streaks}. {\em J. Fluid Mech.\/} {\bf 428},
  29--60.

\bibitem[Antonia {\em et~al.\/}(1988)Antonia, Fulachier, Krishnamoorthy,
  Benabid \& Anselmet]{antonia_fulachier_krishnamoorthy_benabid_anselmet_1988}
{\sc Antonia, R.~A., Fulachier, L., Krishnamoorthy, L.~V., Benabid, T. \&
  Anselmet, F.} 1988 Influence of wall suction on the organized motion in a
  turbulent boundary layer. {\em J. Fluid Mech.\/} {\bf 190}, 217--240.

\bibitem[Brandt {\em et~al.\/}(2004)Brandt, Schlatter \&
  Henningson]{brandt_schlatter_henningson_2004}
{\sc Brandt, L., Schlatter, P. \& Henningson, D.~S.} 2004 Transition in
  boundary layers subject to free-stream turbulence. {\em J. Fluid Mech.\/}
  {\bf 517}, 167--198.

\bibitem[Cherubini {\em et~al.\/}(2011)Cherubini, {De Palma}, Robinet \&
  Bottaro]{cherubini_depalma_robinet_bottaro_2011}
{\sc Cherubini, S., {De Palma}, P., Robinet, J.~C. \& Bottaro, A.} 2011 Edge
  states in a boundary layer. {\em Phys. Fluids\/} {\bf 23}, 051705.

\bibitem[Chevalier {\em et~al.\/}(2007)Chevalier, Schlatter, Lundbladh \&
  Henningson]{chevalier_schlatter_lundbladh_henningson_2007}
{\sc Chevalier, M., Schlatter, P., Lundbladh, A. \& Henningson, D.~S.} 2007 A
  pseudo-spectral solver for incompressible boundary layer flows. {\em Tech.
  Rep.\/} TRITA-MEK 2007:07. KTH Mechanics, Stockholm, Sweden.

\bibitem[Duguet {\em et~al.\/}(2013)Duguet, Monokrousos, Brandt \&
  Henningson]{duguet_monokrousos_brandt_henningson_2013}
{\sc Duguet, Y., Monokrousos, A., Brandt, L. \& Henningson, D.~S.} 2013 Minimal
  transition thresholds in plane {C}ouette flow. {\em Phys. Fluids\/} {\bf 25},
  084103.

\bibitem[Duguet {\em et~al.\/}(2009)Duguet, Schlatter \&
  Henningson]{duguet_schlatter_henningson_2009}
{\sc Duguet, Y., Schlatter, P. \& Henningson, D.~S.} 2009 Localized edge states
  in plane {C}ouette flow. {\em Phys. Fluids\/} {\bf 21}, 111701.

\bibitem[Duguet {\em et~al.\/}(2010{\natexlab{{\em a\/}}})Duguet, Schlatter \&
  Henningson]{duguet_schlatter_henningson_2010}
{\sc Duguet, Y., Schlatter, P. \& Henningson, D.~S.} 2010{\natexlab{{\em a\/}}}
  Formation of turbulent patterns near the onset of transition in plane
  {C}ouette flow. {\em J. Fluid Mech.\/} {\bf 650}, 119--129.

\bibitem[Duguet {\em et~al.\/}(2012)Duguet, Schlatter, Henningson \&
  Eckhardt]{duguet_schlatter_henningson_eckhardt_2012}
{\sc Duguet, Y., Schlatter, P., Henningson, D.~S. \& Eckhardt, B.} 2012
  Self-sustained localized structures in a boundary-layer flow. {\em Phys. Rev.
  Lett.\/} {\bf 108}, 044501.

\bibitem[Duguet {\em et~al.\/}(2010{\natexlab{{\em b\/}}})Duguet, Willis \&
  Kerswell]{duguet_willis_kerswell_2010}
{\sc Duguet, Y., Willis, A.~P. \& Kerswell, R.~R.} 2010{\natexlab{{\em b\/}}}
  Slug genesis in cylindrical pipe flow. {\em J. Fluid Mech.\/} {\bf 663},
  180--208.

\bibitem[Emmons(1951)]{emmons_1951}
{\sc Emmons, H.~W.} 1951 The laminar-turbulent transition in a boundary layer -
  {P}art {I}. {\em J. Aero. Sci.\/} {\bf 18}~(7), 490--498.

\bibitem[Fransson \& Alfredsson(2003)]{fransson_alfredsson_2003}
{\sc Fransson, J. H.~M. \& Alfredsson, P.~H.} 2003 On the disturbance growth in
  an asymptotic suction boundary layer. {\em J. Fluid Mech.\/} {\bf 482},
  51--90.

\bibitem[Griffith \& Meredith(1936)]{griffith_meredith_1936}
{\sc Griffith, A.~A. \& Meredith, F.~W.} 1936 The possible improvement in
  aircraft performance due to the use of boundary layer suction. {\em Tech.
  Rep.\/} 3501. Royal Aircraft Establishment.

\bibitem[Hamilton {\em et~al.\/}(1995)Hamilton, Kim \&
  Waleffe]{hamilton_kim_waleffe_1995}
{\sc Hamilton, J.~M., Kim, J. \& Waleffe, F.} 1995 Regeneration mechanisms of
  near-wall turbulence structures. {\em J. Fluid Mech.\/} {\bf 287}, 317--348.

\bibitem[Hocking(1975)]{hocking_1975}
{\sc Hocking, L.~M.} 1975 Non-linear instability of the asymptotic suction
  velocity profile. {\em Q. J. Mech. Appl. Math.\/} {\bf 28}~(3), 341--353.

\bibitem[Itano \& Toh(2001)]{itano_toh_2001}
{\sc Itano, T. \& Toh, S.} 2001 The dynamics of bursting process in wall
  turbulence. {\em J. Phys. Soc. Jpn.\/} {\bf 70}, 703--716.

\bibitem[Kendall(1998)]{kendall_1998}
{\sc Kendall, J.~M.} 1998 Experiments on boundary-layer receptivity to
  free-stream turbulence. {\textit{AIAA Paper}}. 98-0530.

\bibitem[Kerswell {\em et~al.\/}(2014)Kerswell, Pringle \&
  Willis]{kerswell_pringle_willis_2014}
{\sc Kerswell, R.~R., Pringle, C. C.~T. \& Willis, A.~P.} 2014 An optimization
  approach for analysing nonlinear stability with transition to turbulence in
  fluids as an exemplar. {\em Rep. Prog. Phys.\/} {\bf 77}, 085901.

\bibitem[Khapko {\em et~al.\/}(2014)Khapko, Duguet, Kreilos, Schlatter,
  Eckhardt \&
  Henningson]{khapko_duguet_kreilos_schlatter_eckhardt_henningson_2013}
{\sc Khapko, T., Duguet, Y., Kreilos, T., Schlatter, P., Eckhardt, B. \&
  Henningson, D.~S.} 2014 Complexity of localised coherent structures in a
  boundary-layer flow. {\em Eur. Phys. J. E\/} {\bf 37}~(32).

\bibitem[Khapko {\em et~al.\/}(2013)Khapko, Kreilos, Schlatter, Duguet,
  Eckhardt \&
  Henningson]{khapko_kreilos_schlatter_duguet_eckhardt_henningson_2013}
{\sc Khapko, T., Kreilos, T., Schlatter, P., Duguet, Y., Eckhardt, B. \&
  Henningson, D.~S.} 2013 Localised edge states in the asymptotic suction
  boundary layer. {\em J. Fluid Mech.\/} {\bf 717}, R6.

\bibitem[Khapko {\em et~al.\/}(2016)Khapko, Schlatter, Duguet \&
  Henningson]{khapko_schlatter_duguet_henningson_2015}
{\sc Khapko, T., Schlatter, P., Duguet, Y. \& Henningson, D.~S.} 2016
  Turbulence collapse in a suction boundary layer. {\em J. Fluid Mech.\/} {\bf
  795}, 356--379.

\bibitem[Kreilos {\em et~al.\/}(2016)Kreilos, Khapko, Schlatter, Duguet,
  Henningson \&
  Eckhardt]{kreilos_khapko_schlatter_duguet_henningson_eckhardt_2015}
{\sc Kreilos, T., Khapko, T., Schlatter, P., Duguet, Y., Henningson, D.~S. \&
  Eckhardt, B.} 2016 Bypass transition and spot nucleation in boundary layers.
  {\em Phys. Rev. Fluids\/} ~(under review).

\bibitem[Kreilos {\em et~al.\/}(2013)Kreilos, Veble, Schneider \&
  Eckhardt]{kreilos_veble_schneider_eckhardt_2013}
{\sc Kreilos, T., Veble, G., Schneider, T.~M. \& Eckhardt, B.} 2013 Edge states
  for the turbulence transition in the asymptotic suction boundary layer. {\em
  J. Fluid Mech.\/} {\bf 726}, 100--122.

\bibitem[Matsubara \& Alfredsson(2001)]{matsubara_alfredsson_2001}
{\sc Matsubara, M. \& Alfredsson, P.~H.} 2001 Disturbance growth in boundary
  layers subjected to free-stream turbulence. {\em J. Fluid Mech.\/} {\bf 430},
  149--168.

\bibitem[Mellibovsky {\em et~al.\/}(2009)Mellibovsky, Meseguer, Schneider \&
  Eckhardt]{mellibovsky_meseguer_schneider_eckhardt_2009}
{\sc Mellibovsky, F., Meseguer, A., Schneider, T.~M. \& Eckhardt, B.} 2009
  Transition in localized pipe flow turbulence. {\em Phys. Rev. Lett.\/} {\bf
  103}, 054502.

\bibitem[Saric {\em et~al.\/}(2002)Saric, Reed \&
  Kerschen]{saric_reed_kerschen_2002}
{\sc Saric, W.~S., Reed, H.~L. \& Kerschen, E.~J.} 2002 Boundary-layer
  receptivity to freestream disturbances. {\em Annu. Rev. Fluid Mech.\/} {\bf
  34}, 291--319.

\bibitem[Schlatter {\em et~al.\/}(2008)Schlatter, Brandt, {De Lange} \&
  Henningson]{schlatter_brandt_delange_henningson_2008}
{\sc Schlatter, P., Brandt, L., {De Lange}, H.~C. \& Henningson, D.~S.} 2008 On
  streak breakdown in bypass transition. {\em Phys. Fluids\/} {\bf 20}, 101505.

\bibitem[Schneider {\em et~al.\/}(2007)Schneider, Eckhardt \&
  Yorke]{schneider_eckhardt_yorke_2007}
{\sc Schneider, T.~M., Eckhardt, B. \& Yorke, J.~A.} 2007 Turbulence transition
  and the edge of chaos in pipe flow. {\em Phys. Rev. Lett.\/} {\bf 99},
  034502.

\bibitem[Schneider {\em et~al.\/}(2010)Schneider, Marinc \&
  Eckhardt]{schneider_marinc_eckhardt_2010}
{\sc Schneider, T.~M., Marinc, D. \& Eckhardt, B.} 2010 Localized edge states
  nucleate turbulence in extended plane {C}ouette cells. {\em J. Fluid Mech.\/}
  {\bf 646}, 441--451.

\bibitem[Skufca {\em et~al.\/}(2006)Skufca, Yorke \&
  Eckhardt]{skufca_yorke_eckhardt_2006}
{\sc Skufca, J.~D., Yorke, J.~A. \& Eckhardt, B.} 2006 Edge of chaos in a
  parallel shear flow. {\em Phys. Rev. Lett.\/} {\bf 96}, 174101.

\bibitem[Willis \& Kerswell(2009)]{willis_kerswell_2009}
{\sc Willis, A.~P. \& Kerswell, R.~R.} 2009 Turbulent dynamics of pipe flow
  captured in a reduced model: puff relaminarization and localized `edge'
  states. {\em J. Fluid Mech.\/} {\bf 619}, 213--233.

\bibitem[Yoshioka {\em et~al.\/}(2004)Yoshioka, Fransson \&
  Alfredsson]{yoshioka_fransson_alfredsson_2004}
{\sc Yoshioka, S., Fransson, J. H.~M. \& Alfredsson, P.~H.} 2004 Free stream
  turbulence induced disturbances in boundary layers with wall suction. {\em
  Phys. Fluids\/} {\bf 16}~(10), 3530--3539.

\bibitem[Zaki \& Durbin(2005)]{zaki_durbin_2005}
{\sc Zaki, T.~A. \& Durbin, P.~A.} 2005 Mode interaction and the bypass route
  to transition. {\em J. Fluid Mech.\/} {\bf 531}, 85--111.

\bibitem[Zammert \& Eckhardt(2014)]{zammert_eckhardt_2014}
{\sc Zammert, S. \& Eckhardt, B.} 2014 Streamwise and doubly-localised periodic
  orbits in plane {P}oiseuille flow. {\em J. Fluid Mech.\/} {\bf 761},
  348--359.

\end{thebibliography}
\end{document}